\documentclass[
onecolumn
]{aastex631}
\usepackage{graphics} 
\usepackage{epsfig} 
\usepackage{mathptmx} 
\usepackage{times} 
\usepackage{amsmath} 
\usepackage{amssymb}  
\usepackage{xcolor}
\usepackage{subfigure}
\usepackage{bm}
\usepackage{hyperref}

\graphicspath{{./}}

\accepted{April 13, 2025 }

\submitjournal{ApJL}

\shorttitle{Transfer of Entropy between the Magnetic Field and SEPs during an ICME}
\shortauthors{Cuesta et al.}

\begin{document}

\title{Transfer of Entropy between the Magnetic Field and Solar Energetic Particles during an Interplanetary Coronal Mass Ejection}

\author[0000-0002-7341-2992]{M. E. Cuesta}
\affiliation{Department of Astrophysical Sciences,
Princeton University, Princeton, NJ 08544, USA}
\correspondingauthor{Manuel Enrique Cuesta}
\email{mecuesta@princeton.edu}

\author[0000-0002-7655-6019]{G. Livadiotis}
\affiliation{Department of Astrophysical Sciences, Princeton University, Princeton, NJ 08544, USA}

\author[0000-0001-6160-1158]{D. J. McComas}
\affiliation{Department of Astrophysical Sciences, Princeton University, Princeton, NJ 08544, USA}


\author[0000-0003-0412-1064]{L. Y. Khoo}
\affiliation{Department of Astrophysical Sciences, Princeton University, Princeton, NJ 08544, USA}

\author[0000-0001-7952-8032]{H. A. Farooki}
\affiliation{Department of Astrophysical Sciences, Princeton University, Princeton, NJ 08544, USA}


\author[0000-0002-6962-0959]{R. Bandyopadhyay}
\affiliation{Department of Astrophysical Sciences, Princeton University, Princeton, NJ 08544, USA}

\author[0000-0002-1989-3596]{S. D. Bale}
\affiliation{Physics Department, University of California, Berkeley, CA, 94720, USA}
\affiliation{Space Sciences Laboratory, University of California, Berkeley, CA 94720, USA}

\begin{abstract}
    Thermodynamics of solar wind bulk plasma have been routinely measured and quantified, unlike those of solar energetic particles (SEPs), whose thermodynamic properties have remained elusive until recently.
    The thermodynamic kappa (\(\kappa_{\rm EP}\)) that parameterizes the statistical distribution of SEP kinetic energy contains information regarding the population's level of correlation and effective degrees of freedom (\({\rm d_{eff}}\)).
    At the same time, the intermittent kappa (\(\kappa_{\Delta B}\)) that parameterizes the statistical distribution of magnetic field increments contains information about the correlation and \({\rm d_{eff}}\) involved in magnetic field fluctuations.
    Correlations between particles can be affected by magnetic field fluctuations, leading to a relationship between \(\kappa_{\rm EP}\) and \(\kappa_{\Delta B}\).
    In this paper, we examine the relationship of \({\rm d_{eff}}\) and entropy between energetic particles and the magnetic field via the spatial variation of their corresponding parameter kappa values.
    We compare directly the values of \(\kappa_{\rm EP}\) and \(\kappa_{\Delta B}\) using Parker Solar Probe IS\(\odot\)IS and FIELDS measurements during an SEP event associated with an interplanetary coronal mass ejection (ICME).
    Remarkably, we find that \(\kappa_{\rm EP}\) and \(\kappa_{\Delta B}\) are anti-correlated via a linear relationship throughout the passing of the ICME, indicating a proportional exchange of \({\rm d_{eff}}\) from the magnetic field to energetic particles, i.e., \(\kappa_{\Delta B} \sim (-0.15 \pm 0.03)\kappa_{\rm EP}\), interpreted as an effective coupling ratio.
    This finding is crucial for improving our understanding of ICMEs and suggests that they help to produce an environment that enables the transfer of entropy from the magnetic field to energetic particles due to changes in intermittency of the magnetic field.
    \newline
\end{abstract}



\section{Introduction} \label{sec:intro}

Interplanetary space is permeated by solar wind plasma and energetic particles of both solar and interstellar origin.
Protons comprise the bulk of the solar wind, flowing along the interplanetary magnetic field lines.
As a result, long-range particle correlations induced by the magnetic field persist in the solar wind, as in other space plasmas, usually over scales as large as the energy-containing scale (or the correlation scale, \(\sim 10^6\)~km at 1~au; \citet{MatthaeusEA1982}).
Due to these long-range correlations between particles and fields, classical thermodynamic frameworks---such as the Maxwellian canonical velocity distribution \citep{Maxwell1860} and \citet{Boltzmann1866} -- \citet{Gibbs1902elementary_book} (BG) entropy---fail to describe the full particle velocity/energy distributions observed in space plasmas.
In the solar wind, it is observed that the distribution of the bulk (energies \(\sim 10\)~eV) is well characterized by a Maxwellian core \citep{NeugebauerSnyder1966JGR_Mariner2_SolarWindObservations,Hundhausen1968SSRv_DirectSWobservations,FeldmanEA1975JGR_SWelectrons}. 
At the same time, high-energy tails are observed in the distribution with far more energetic particles than there would be in a purely Maxwellian distribution \citep{SarrisEA1981GRL_EPs_nonMaxwellian}.
Therefore, a generalized framework of thermodynamics is required to characterize the full velocity distribution \citep{LivadiotisMcComas2013SSRv_KappaDist_SpSc,LivadiotisMcComas2023ApJ_ConnectionHeatingPolytropicIndex,Livadiotis2017kappa_book}.

Kappa distributions and their associated statistical mechanics and thermodynamics have been applied successfully to space plasmas throughout the heliosphere.
For instance, solar wind plasma parameters have frequently been determined through modeling observed velocity distributions as kappa distributions \citep{CollierEA1996GeoRL_HeavyDensities_SW,MaksimovicEA1997AAP_KappaCoronaSW,MaksimovicEA1997GeoRL_ElectronKappa_Ulysses,PierrardEA1999JGR_ElectronVDFs,marsch2006kinetic,Zouganelis2008JGRA_SuprathermalParams_QTNkappa,StverakEA2009JGRA_RadEvol_ElectPops,LivadiotisMcComas2010ApJ_SpacePlasmaNonEquil,Yoon2012PhPl_ElectronKappa_LangmuirTurb,Yoon2014JGRA_ElectronKappa_QTN}.
Although thermodynamic parameters (kappa, temperature, and density) of the solar wind are commonly derived from observations throughout the heliosphere, these quantities have remained elusive for solar energetic particles (SEPs).

\begin{figure}[ht]
    \centering
    \includegraphics[width=0.6\linewidth]{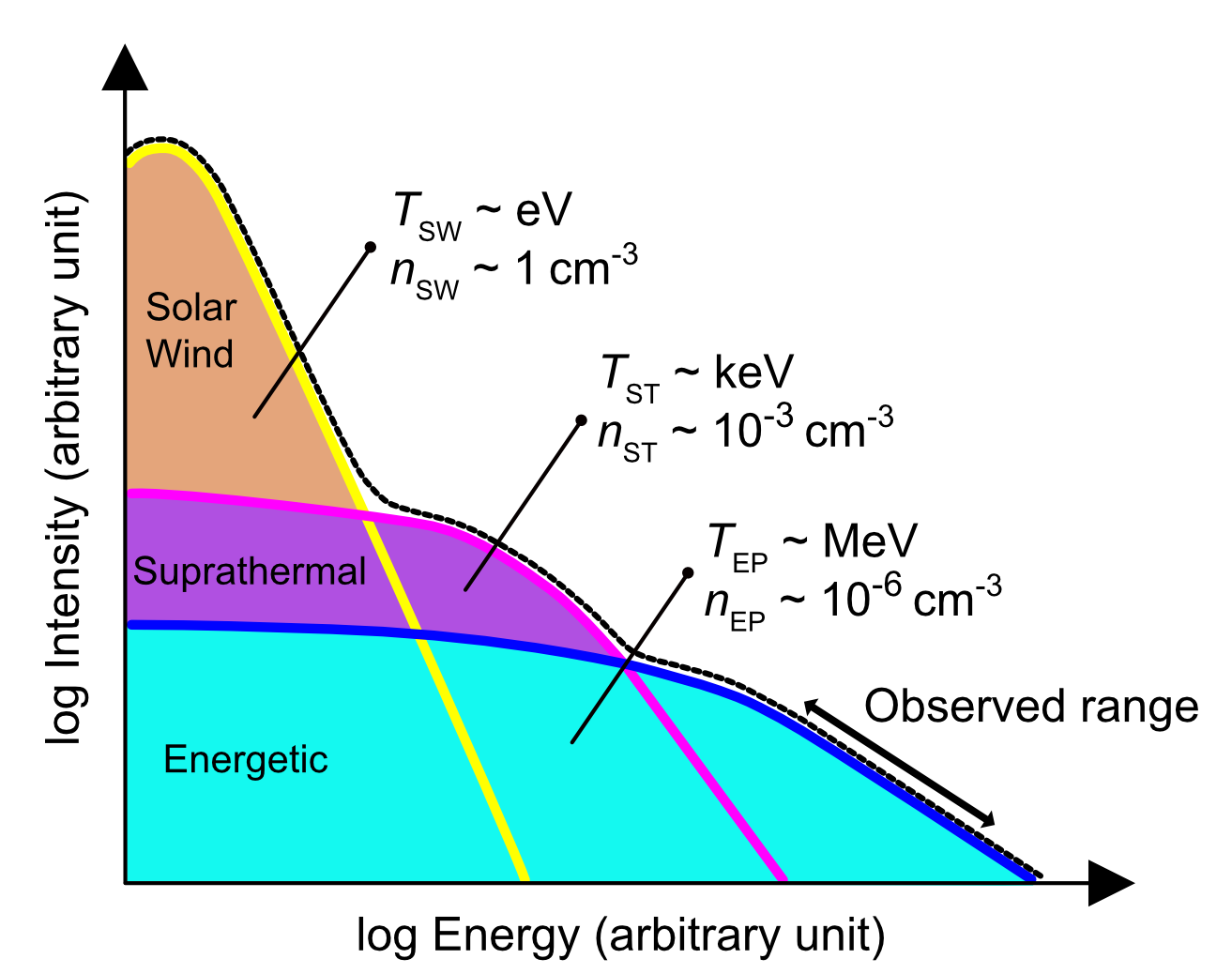}
    \caption{The distributions of the solar wind (SW), suprathermal (ST), and energetic particle (EP) populations as a function of energy. This illustration was taken from \citet{CuestaEA2024ApJ_EPthermoKappaObservations}.}
    \label{fig:populations}
\end{figure}

SEPs include protons, electrons, and alpha particles with energies much greater than the bulk plasma.
Suprathermal particles, a subset of energetic particles, have energies generally in the range from \(\sim 10\)~keV to 1~MeV per nucleon \citep{MasonGloeckler2012SSRv_suprathermals}, which connect the Maxwellian- or kappa-distributed solar wind particles of higher energy with SEPs.
In Figure \ref{fig:populations}, we illustrate how the intensity distributions of the solar wind, suprathermal, and energetic particle populations compose the overall observed distribution.
Suprathermal particles can act as a seed population for further energization by acceleration mechanisms, such as diffusive shock acceleration (DSA).
DSA is believed to be the main process responsible for accelerating energetic particles during interplanetary shocks \citep{Fisk1971JGR_CosRayIntensity_IPshock,AxfordEA1977ICRC_CosRayAccel_Shock,BlandfordOstriker1978ApJL_ParticleAccel_Shock,Bell1978MNRAS_ShockAccel_CosRay,GoslingEA1979AIPC_IonAccel_EarthBowShock,Lee1983JGR_IonAccel_IPshock,Lee1997GMS_CMEshock_transport}.
Particle acceleration due to magnetic flux rope deformation and merging has also been regarded as an efficient process of particle energization \citep{DrakeEA2006NatAcceleration,OkaEA2010ApJ_ElectronAccel_Coalescence, DrakeEA2013ApJL_EnergeticParticleSpectra_Reconnection,PhanEA2024ApJL_MagReconnection_HCS_FluxRopeMerging,DesaiEA2024arXiv_HCSMagReconnection_ProtonEnergization}.

Recently, a technique developed for kappa-tail fitting was applied to the distribution of energetic proton intensities as a function of energy \citep{Livadiotis2007_FittingGeneralMethods,LivadiotisEA2024ApJ_EPthermoKappaTechnique,CuestaEA2024ApJ_EPthermoKappaObservations}.
This method estimates the thermodynamic parameters---kappa, density, and temperature---that parameterize the kappa distribution underlying the observed intensity-energy spectrum.
Besides the thermodynamic kappa of energetic protons (\(\kappa_{\rm EP}\)), one can derive a parameter kappa for the magnetic field (\(\kappa_{\Delta B}\)) based on magnetic increments.
Magnetic increments are important in solar wind turbulence for measuring the turbulent energy cascade \citep{Kolmogorov1941,Kolmogorov1962,Obukhov1962}, and can be used to detect strong discontinuities, either directional or in magnitude, in the magnetic field \citep{GrecoEA2008,GrecoEA2012}.
The distributions of magnetic increments have been observed to follow a kappa distribution \citep{GrecoEA2018}, implying that it is possible to derive \(\kappa_{\Delta B}\).
In this paper, we use two methods to derive \(\kappa_{\Delta B}\), which we find to provide consistent results. We compare the values of \(\kappa_{\Delta B}\) derived from one of these methods to \(\kappa_{\rm EP}\) over the passing of an interplanetary coronal mass ejection (ICME) using Parker Solar Probe \citep[PSP;][]{Fox2016SSRv_PSP} observations.
We also compare the block entropy of the magnetic field magnitude \citep{Shannon1948IEEE_InformationEntropy, Grassberger1989IEEE_BlockEntropy,Livadiotis2008PhyA_BlockEntropy} to the thermodynamic entropy of energetic protons, finding a similar relationship to the relationship between \(\kappa_{\Delta B}\) and \(\kappa_{\rm EP}\).

The paper is organized as follows.
In Section \ref{sec:data}, we describe the data we use in our analysis.
We describe two methods for acquiring values of \(\kappa_{\Delta B}\) in Sections \ref{subsubsec:mag_kappa_fit} and \ref{subsubsec:mag_kappa_mean} (values of \(\kappa_{\rm EP}\) are taken from \citet{CuestaEA2024ApJ_EPthermoKappaObservations} and described in Section \ref{subsec:EP_kappa}).
Also, we describe a type of information entropy in Section \ref{subsec:block_entropy} derived from the magnitude of the magnetic field.
In Section \ref{sec:results}, we show our results, where we compare the thermodynamic and intermittent kappa and their respective entropies, which are discussed in Section \ref{sec:discussion} and summarized in Section \ref{sec:summary}.

\section{Data Description} \label{sec:data}

We utilize energetic proton intensities measured at 1~minute cadence by the ``sunward''-facing side of the EPI-Hi high energy telescope (HET-A), which is part of the Integrated Science Investigation of the Sun \citep[IS\(\odot\)IS; ][]{McComasEA2016SSRv_ISOIS} instrument suite onboard PSP.
HET is mounted with its sunward-facing boresight pointed along a line 20 degrees from the Sun-spacecraft line, which is 25 degrees less than a nominal Parker spiral direction for a 400~km/s solar wind speed at 1~au.
Energetic protons (10 -- 60~MeV) are observed during the SEP event that occurred on 2022 February 15
at a radial distance of \(\sim 0.35~{\rm au}\)
, with an interplanetary shock crossing at around 2022 February 16 07:25 \citep{PalmerioEA2024ApJ_CME_BepiPSP_2022feb15,KhooEA2024ApJ_CME_BepiPSP_2022feb15}.
The range of energies are selected to avoid any potential sensitivity resolution issues within the full range of provided energy bins.
We also use magnetic field measurements, which are measured by the PSP/FIELDS \citep{BaleEA2016} magnetometer \citep{BowenEA2020JGR_FIELDSmagnetometer}.
These measurements are downsampled from 4-samples-per-cycle cadence to 1-second samples.
All data are publicly available at NASA's Space Physics Data Facility, which can be accessed at \url{https://cdaweb.gsfc.nasa.gov/}.

\section{Methodology} \label{sec:methods}

\subsection{Intermittent Kappa of Magnetic Field} \label{subsec:Mag_kappa}

Magnetic field increments are useful in determining the strength of magnetic discontinuities over a given increment scale.
The partial variance of increments (PVI) method, which utilizes magnetic increments, describes the roughness of the magnetic field \citep{GrecoEA2008}.
This roughness of the field is determined relative to the background level of absolute increments over a period longer than the increment scale \citep{GrecoEA2008,GrecoEA2012,GrecoEA2018,ServidioEA2011JGRA_PVIaveraging_Reconnection}.
Here, we examine the increments of the individual components of the magnetic field (\(B_i\) for \(i \in \{{\rm R,T,N}\}\) are the radial, tangential, and normal directions in the heliocentric inertial frame; \citet{FranzHarperEA2002PSS_HelioCoords}).
Magnetic increments are sensitive to discontinuities over magnitude and direction.
A magnetic increment is defined as:

\begin{equation}\label{eq:increment}
    \Delta \bm{B}(t,\tau_{\rm lag}) = \bm{B}(t+\tau_{\rm lag}) - \bm{B}(t)\text{,}
\end{equation}
where \(\tau_{\rm lag}\) is the increment time scale over time \(t\) and \(\bm{B}\) is the magnetic field vector with components \(B_i\) as defined above.
We follow the considerations of \citet{ServidioEA2011JGRA_PVIaveraging_Reconnection} to determine the value of \(\tau_{\rm lag}\) with respect to PVI analysis.
We choose an increment scale within the inertial range that is much less than the correlation, or energy-containing, scale.
The inertial range is defined as the range of scales between the correlation and dissipation, or inner, scale.
First, we must estimate the values of the correlation and dissipation scale, the latter which is commonly associated with the ion inertial scale (\({\rm d_i}\)).
To estimate \({\rm d_i}\) within 1~au, we extrapolate the observed radial trend of \({\rm d_i}\) \citep[\({\rm d_i}\sim 80~{\rm km}\) at 1~au;][]{CuestaEA2022ApJS_RadialIntermittency}
to the radial distance where PSP observed the SEP event under analysis. 
Assuming a 450~km/s solar wind speed, the ion inertial scale is estimated at \(\sim 28\)~km, or \(\sim 0.06\)~s at the radial distance of \(\sim 0.35~{\rm au}\).
Furthermore, \citet{CuestaEA2022ApJS_RadialIntermittency,CuestaEA2022ApJL_Isotropization} give radial trends of the correlation scale, which scales as the square-root of heliocentric distance, on average.
With an average scale of 45~minutes at 1~au \citep{MatthaeusEA1982}, we find that the correlation scale at 0.35~au is \(\sim 27\)~minutes.
As a result, we set \(\tau_{\rm lag}=10~{\rm s}\) for the present analysis, which is \(\sim 167~{\rm d_i}\) on average and is much less than the correlation scale.

The distribution of magnetic increments has been previously examined and characterized as having super-Gaussian tails \citep{GrecoEA2018}. It is suggested that kappa distributions are useful in describing the distribution of magnetic increments, as was demonstrated in prior observations of the interplanetary magnetic field \citep{BurlagaVinas2005PhyA_Voyager_QnonExtensive}, Earth's magnetosphere \citep{PollockEA2018JASTP_KappaMagElectron_EarthMagnetosphere}, and in the heliopause (Khoo et al. 2025, in prep.).
The kappa distribution formula that provides a statistical description of a vector \(\bm{x}\) is given by \citep{LivadiotisMcComas2013SSRv_KappaDist_SpSc}:

\begin{equation}\label{eq:kappa_vector}
    P(\bm{x}) = (2\pi \kappa_0\sigma_x^2)^{(-D/2)} \cdot \frac{\Gamma(\kappa_0+1+D/2)}{\Gamma(\kappa_0+1)} \cdot \left[ 1+\frac{D}{\kappa_0} \frac{(\bm{x}-\mu_x)^2}{2\sigma_x^2} \right]^{(-\kappa_0-1-D/2)}\text{,}
\end{equation}
where the normalization is for each component of \(\bm{x}\) such that \(x_i \in (-\infty,+\infty)\); \(D\) denotes the dimensionality of the magnetic component whose increments are examined (e.g., \(D=1\) for each vector component, \(D=2\) for planar projections, \(D=3\) for the whole vector magnitude) and \(\kappa_0=\kappa-D/2\) is the invariant kappa parameter \citep[e.g.,][]{LivadiotisMcComas2011ApJ_InvariantKappaDistribution,Livadiotis2015Entrp_Kappa_DegreeOfFreedom}.
\(\bm{\mu}_x\) is the mean vector of the distributed \(x\)-values, \(\sigma_x\) is the respective standard deviation of each one-dimensional component of \(\bm{x}\) values, and \(\Gamma(a)\) is the Gamma function of some argument \(a\).
We set \(x=\Delta B_i\) (all meanless, \(\mu_x=0\), and one dimensional, \(D=1\)) in Eq. \eqref{eq:kappa_vector}, which gives:

\begin{equation}\label{eq:increment_kappa}
    P(|\Delta B_i|) = 2\cdot P(\Delta B_i) = 2(2\pi\sigma^2_{\Delta B_i}(\kappa_{\Delta B_i}-3/2))^{(-1/2)} \cdot \frac{\Gamma(\kappa_{\Delta B_i})}{\Gamma(\kappa_{\Delta B_i}-1/2)} \cdot \left[ 1+\frac{1}{2\kappa_{\Delta B_i}-3} \cdot \frac{(\Delta B_i)^2}{\sigma^2_{\Delta B_i}}\right]^{-\kappa_{\Delta B_i}}\text{,}
\end{equation}
resulting in the kappa distribution describing the magnetic increments of any single vector component of the magnetic field, where \(\kappa_{\Delta B_i} \equiv \kappa_0 + 3/2\) is the three-dimensional intermittent kappa for that corresponding component of the magnetic increment (similarly defined with the thermodynamic kappa, i.e., \(\kappa_{\rm EP}\equiv \kappa_0 + 3/2\)).
We have included a factor of two since the normalization takes place for \(\Delta B_i\in[0,\infty]\).

\subsection{Finding the Intermittent Kappa} \label{subsec:Finding_kappa}

To find the value of \(\kappa_{\Delta B_i}\) for an arbitrary interval, we use two techniques: (1) kappa fitting and (2) kappa moments.
Then, the overall intermittent kappa (\(\kappa_{\Delta B}\)) is found by taking an arithmetic average of all \(\kappa_{\Delta B_i}\), given as:

\begin{equation}\label{eq:overall_mag_kappa}
    \kappa_{\Delta B} = \frac{1}{3} \left(\kappa_{\Delta B_{\rm R}}+\kappa_{\Delta B_{\rm T}}+\kappa_{\Delta B_{\rm N}} \right)\text{.}
\end{equation}
Next, we describe the two techniques for finding the kappa value of the distributions of magnetic increments.

\subsubsection{Fitting Kappa Distributions} \label{subsubsec:mag_kappa_fit}

We model the logarithm (base 10) of the probability density function (PDF) of \(\Delta B_i\) according to the logarithm of Eq. \eqref{eq:increment_kappa}:

\begin{subequations}
\begin{align}\label{eq:fit_kappa}
    \log \left[ P(\Delta B_i) \right] = A_1 - \kappa_{\Delta B_i} \cdot \log \left[ 1 + A_2 \cdot (\Delta B_i)^2 \right]
\end{align}
where

\begin{align}
    10^{A_1} = 2\left[ 2\pi \sigma^2_{\Delta B_i} (\kappa_{\Delta B_i} -3/2) \right]^{-1/2} \cdot \frac{\Gamma(\kappa_{\Delta B_i})}{\Gamma(\kappa_{\Delta B_i - 1/2})}
\end{align}
and

\begin{align}
    \frac{1}{A_2} = (2\kappa_{\Delta B_i}-3)\sigma^2_{\Delta B_i}\text{.}
\end{align}
\end{subequations}
This fitting procedure yields a fitted intermittent kappa for each component of the magnetic increments, with a corresponding fitting error.
Then, we take an arithmetic average to find the overall intermittent kappa, according to Eq. \eqref{eq:overall_mag_kappa}, resulting in values of \(\kappa^{\rm (fit)}_{\Delta B}\).

\subsubsection{Moments of Kappa Distributions} \label{subsubsec:mag_kappa_mean}

In addition to the kappa fitting method to find values of \(\kappa^{\rm (fit)}_{\Delta B_i}\), we derive \(\kappa_{\Delta B_i}\) another way that makes use of ``\(\alpha\)'' power-means \citep{LivadiotisMcComas2012ApJ_KappaThermoProcesses} or moments of the distribution, \({\rm M}_i(\alpha)\), such that:

\begin{equation}\label{eq:powermean}
     {\rm M}_i(\alpha) = \langle \left(|\Delta B_i|^2 \right)^\alpha \rangle^{1/\alpha} = \left[ \kappa_0^\alpha \frac{\Gamma(\kappa_0+1-\alpha)\cdot \Gamma(\alpha+1/2)}{\Gamma(\kappa_0+1)\cdot \Gamma(1/2)} \right]^{1/\alpha}\text{,}
 \end{equation}
 where the exponent gives the order of the moment, \(0\leq\alpha\leq1\) \citep{Livadiotis2017kappa_book,Livadiotis2019AIPA_CollisionFrequency_MeanFreePath_Kappa}.
 The ratio of two different moments can be used to find an expression that is dimensionless with \(\kappa_{\Delta B_i}\) as the only unknown variable.
 Examining the power-mean for \(\alpha=1\) gives:
 
\begin{equation}\label{eq:powermean_1}
    {\rm M}_i(\alpha=1) = \langle \left(|\Delta B_i|^2 \right) \rangle = \kappa_0 \frac{\Gamma(\kappa_0)\cdot \Gamma(3/2)}{\Gamma(\kappa_0+1)\cdot\Gamma(1/2)} = \frac{1}{2}\text{,}
\end{equation}
and for \(\alpha=1/2\):

\begin{equation}\label{eq:powermean_half}
    {\rm M}_i(\alpha=1/2) = \langle |\Delta B_i| \rangle^{2} = \frac{\kappa_0}{\pi} \left( \frac{\Gamma(\kappa_0+1/2)}{\Gamma(\kappa_0+1)} \right)^2\text{.}
\end{equation}
As a result, the ratio (\({\rm M}_i(1/2)/{\rm M}_i(1)\) gives:

\begin{equation}\label{eq:powermean_ratio}
    \frac{{\rm M}_i(1/2)}{{\rm M}_i(1)} = \frac{\langle |\Delta B_i| \rangle^{2}}{\langle |\Delta B_i|^2 \rangle} = \frac{2\kappa_0}{\pi} \left( \frac{\Gamma(\kappa_0+1/2)}{\Gamma(\kappa_0+1)} \right)^2 \text{.}
\end{equation}
After finding this ratio empirically from the data, we solve Eq. \eqref{eq:powermean_ratio} for \(\kappa^{\rm (m)}_{\Delta B_i}=\kappa_0+3/2\) for each component of the magnetic field increments and take an arithmetic average to find the overall intermittent kappa via Eq. \eqref{eq:overall_mag_kappa}, resulting in values of \(\kappa^{\rm (m)}_{\Delta B_i}\).
To quantify an error associated with this method, we use bootstrapping \citep{Efron1994Book_BootstrapMethod}. 
That is, we repeat the procedure 100 times on random samples of the original magnetic increments. The elements of each random set are selected uniformly from the original set allowing for duplicates.
From these 100 new sets with length equal to the original dataset, we get 100 \(\kappa^{\rm (m)}_{\Delta B_i}\) values, and we take their standard deviation as the uncertainty of \(\kappa^{\rm (m)}_{\Delta B_i}\) calculated using the original set.

For either method of deriving \(\kappa^{\rm (fit)}_{\Delta B}\) or \(\kappa^{\rm (m)}_{\Delta B}\), we use an interval duration of 3~hours shifted by every second (a rolling 3-hour window), yielding possibly over 10,000 increment values per magnetic field component per interval.
A single value of \(\kappa^{\rm (fit)}_{\Delta B}\) and \(\kappa^{\rm (m)}_{\Delta B}\) is derived from one 3-hour interval, a duration based on considerations involving PVI analyses.
\citet{ServidioEA2011JGRA_PVIaveraging_Reconnection} found that the PVI method was most efficient in detecting coherent structures when a sufficiently large background duration was used relative to the increment scale.
Therefore, the choice of 3~hours satisfies this condition.
Additionally, choosing a 3-hour window allowed sufficient sampling statistics for constructing a well-defined PDF for fitting and computing its moments.

\begin{figure}[ht]
    \centering
    \includegraphics[width=\linewidth]{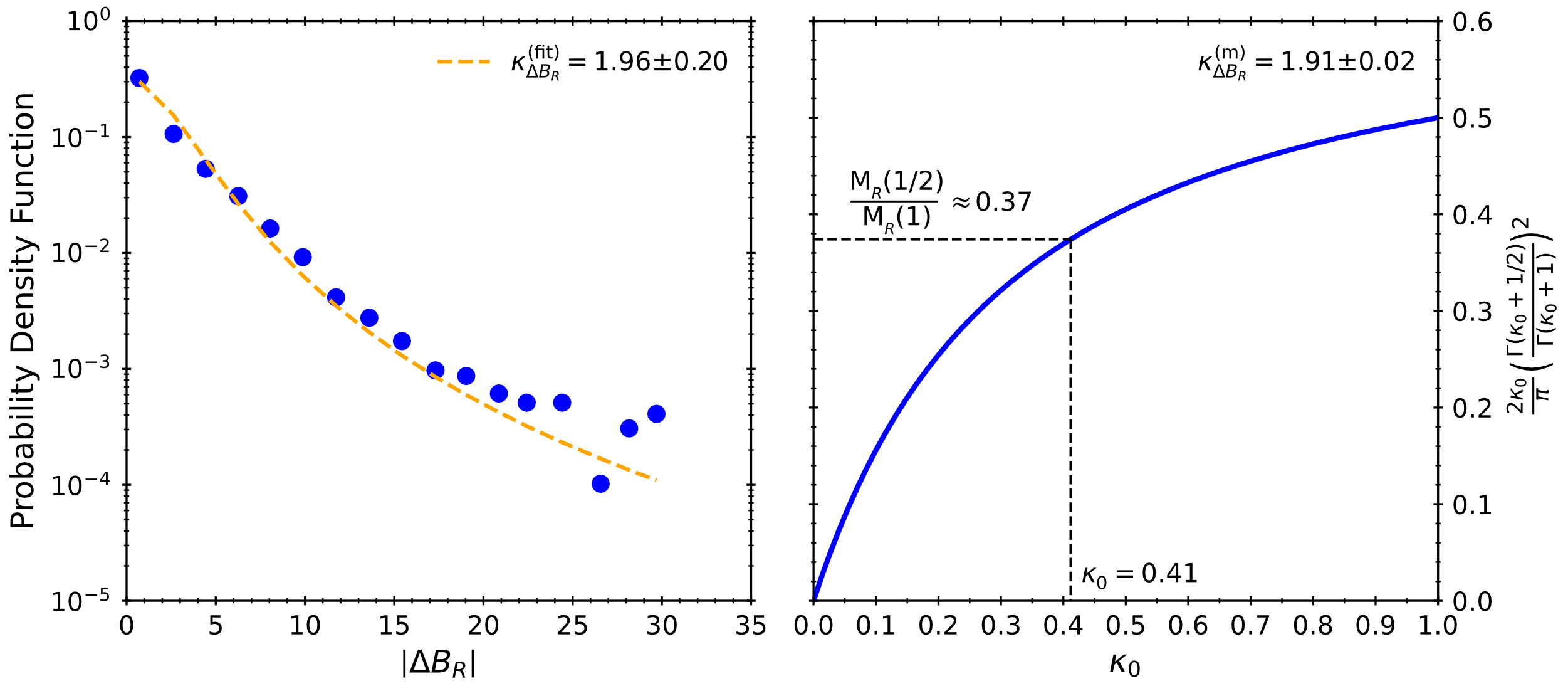}
    \caption{(Left panel): The probability density function of unsigned radial magnetic increments with the best-fit (Eq. \eqref{eq:fit_kappa}) yielding \(\kappa^{\rm (fit)}_{\Delta B_{\rm R}}=1.96 \pm 0.20\). (Right panel): The analytic expression expected for possible ratios of the moments as a function of \(\kappa_0\), yielding \(\kappa^{\rm (m)}_{\Delta B_{\rm R}}=1.91 \pm 0.02\) when compared to the empirically computed ratio from the data (Eq. \eqref{eq:powermean_ratio}). These values of \(\kappa_{\Delta B_{\rm R}}\) correspond to the 3-hour interval centered about 2022-02-16 02:00:00.}
    \label{fig:SampleMethod}
\end{figure}

In Figure \ref{fig:SampleMethod}, we show the derivation of \(\kappa^{\rm (fit)}_{\Delta B_{\rm R}}\) and \(\kappa^{\rm (m)}_{\Delta B_{\rm R}}\) using both methods from a single 3-hour interval centered about 2022-02-16 02:00:00.
Shown in the left panel, we fit Eq. \eqref{eq:fit_kappa} to the PDF of \(|\Delta B_{\rm R}|\), yielding a fitted parameter \(\kappa^{\rm (fit)}_{\Delta B_{\rm R}}=1.96 \pm 0.20\).
On the right panel of Figure \ref{fig:SampleMethod}, we compute the ratio of moments and solve for \(\kappa^{\rm (m)}_{\Delta B_{\rm R}}\) using Eq. \eqref{eq:powermean_ratio}, yielding \(\kappa^{\rm (m)}_{\Delta B_{\rm R}}=1.91 \pm 0.02\), which is consistent with \(\kappa^{\rm (fit)}_{\Delta B_{\rm R}}\) and its fitting error.
For the analysis below, we repeat this step for the normal and tangential components of the magnetic increments and take an arithmetic average (Eq. \eqref{eq:overall_mag_kappa}) to find \(\kappa_{\Delta B}\) as a function of time at a 1-second resolution.

\newpage

\subsection{Thermodynamic Kappa of Energetic Protons} \label{subsec:EP_kappa}

The thermodynamic kappa for energetic particles (\(\kappa_{\rm EP}\)) is a fundamental measure required to assess their thermodynamic behavior from a statistical mechanical perspective.
This is achieved by applying a thermodynamic framework connected to kappa distributions \citep{Livadiotis2007_FittingGeneralMethods,LivadiotisMcComas2010ApJ_SpacePlasmaNonEquil} to observations of SEPs.
Due to the nature of the full velocity distribution observed in interplanetary space, SEPs can only be uniquely observed at high energies far from the solar wind bulk energies (see Figure \ref{fig:populations}).
Nonetheless, the kappa-tail technique \citep{Livadiotis2007_FittingGeneralMethods,Livadiotis2017kappa_book} applied to energetic particle spectra \citep{LivadiotisEA2024ApJ_EPthermoKappaTechnique,CuestaEA2024ApJ_EPthermoKappaObservations} allows us to gain information about the assumed kappa distribution that extends to energies beyond our observed range.
Furthermore, \cite{CuestaEA2024ApJ_EPthermoKappaObservations} applied these procedures to distributions of energetic protons observed from PSP during the 2022 February 15 SEP event, which assumes that the observations over a sufficiently small interval are in a stationary state (small changes over time).
Significant transitions in the observed distributions may weaken the assumption of the observations originating from a single source, or a single kappa distribution.
As a result, \cite{CuestaEA2024ApJ_EPthermoKappaObservations} also provided a way to determine the procedures' applicability in the presence of nonstationarities.

In this paper we use these methods provided by \cite{LivadiotisEA2024ApJ_EPthermoKappaTechnique} and \cite{CuestaEA2024ApJ_EPthermoKappaObservations}, using only values that are deemed optimal \citep[belonging to a stationary period that was confidently determined; see Section 2 of][]{CuestaEA2024ApJ_EPthermoKappaObservations}.
The thermodynamic parameter kappa of energetic protons (\(\kappa_{\rm EP}\)) is derived by fitting the intensity (\(j\)) spectra as a function of energy \(E\), assuming the spectrum is nearly a power-law at energies much larger than the bulk temperature.
As a result, the spectral index (\(\gamma\)) is exactly equal to \(\kappa_{\rm EP}\), for spectrum \(j\sim E^{-\gamma}\) with \(E \gg \kappa_0 k_{\rm B} T\), where \(k_{\rm B}\) is the Boltzmann constant and \(T\) is the temperature of the particle population.
The kappa-tail technique is applied to observations of energetic protons with energies ranging \(\sim\)~10 to 60~MeV \citep{CuestaEA2024ApJ_EPthermoKappaObservations},
which are selected to avoid any potential sensitivity resolution issues at the boundaries of HET's observed energies.
We compare the values of \(\kappa_{\Delta B}\) and \(\kappa_{\rm EP}\) that are found to be interconnected during the passing of an interplanetary coronal mass ejection.

\subsection{Block Entropy of Complexity of the Magnetic field} \label{subsec:block_entropy}

Although entropy gain and loss may be inferred through a corresponding gain and loss in the parameter kappa, here we also compute the block entropy of complexity of the magnetic field magnitude.
Block entropy, in this context, is a measure of the randomness of a sequence, or block, over time \citep{Grassberger1989IEEE_BlockEntropy}. 
The length of these blocks (\(\ell\)) is uniform, dividing the duration of the interval into a \(N\) number of blocks. 
The total possible number of sequences form a probability basis from which the block entropy can be derived.
In this study, we only wish to use the block entropy as a proxy to describe the spatial evolution of the entropy of the magnetic field over the ICME.
The magnitude of the magnetic field encapsulates the information of its components.
Therefore, we choose to examine the magnitude instead of calculating the block entropy for each magnetic field component.
In the future, we plan to investigate the block entropy of the individual components relative to the magnitude in further detail.

Here, we derive the block entropy of an arbitrary interval as follows.
A global median of the magnetic field magnitude is computed, which is used to redefine the time series as a series of zeros and ones: one if the measurement lies above the median and zero if the measurement lies below the median \citep{Livadiotis2008PhyA_BlockEntropy}. 
Next, we set the block length to be four seconds, or \(\ell=4\).
For each non-overlapping block of four points, we tally the total number of each possible sequence, which is equivalent to \(2^{\ell}=16\) combinations.
As a result, the probability of sequence \(s_i\) occurring is:

\begin{equation} \label{eq:block_prob}
    P(s_i,\ell) = \frac{N_{s_i}}{\sum\limits^{2^\ell}_{i=1}N_{s_i}}= 
    \frac{N_{s_i}}{N}\text{,}
\end{equation}
where \(N_{s_i}\) is the number of occurrences for the \(i\)-th possible sequence. 
The block entropy is then defined as:

\begin{equation}\label{eq:block_entropy}
    S_B(\ell) = -\sum\limits_{i=1}^{2^\ell} P(s_i,\ell) \log_2 \left[P(s_i,\ell)\right] \text{,}
\end{equation}
which follows the formalism of Shannon’s information entropy \citep{Shannon1948IEEE_InformationEntropy}. 
To find an error estimation on the value of \(S_B(4)\), we also find the Block entropy for \(\ell\) equal to three and five. 
We then represent the error of \(S_B(4)/{\rm min}[S_B(4)]\) as \(\pm(|S_B(5)-S_B(3))/2{\rm min}[S_B(4)]\).
Below, we examine this entropy-like quantity of the magnetic field as a function of the thermodynamic entropy of energetic protons, which is connected to the thermodynamic kappa.

\section{Results} \label{sec:results}

\begin{figure}[ht]
    \centering
    \includegraphics[width=\linewidth]{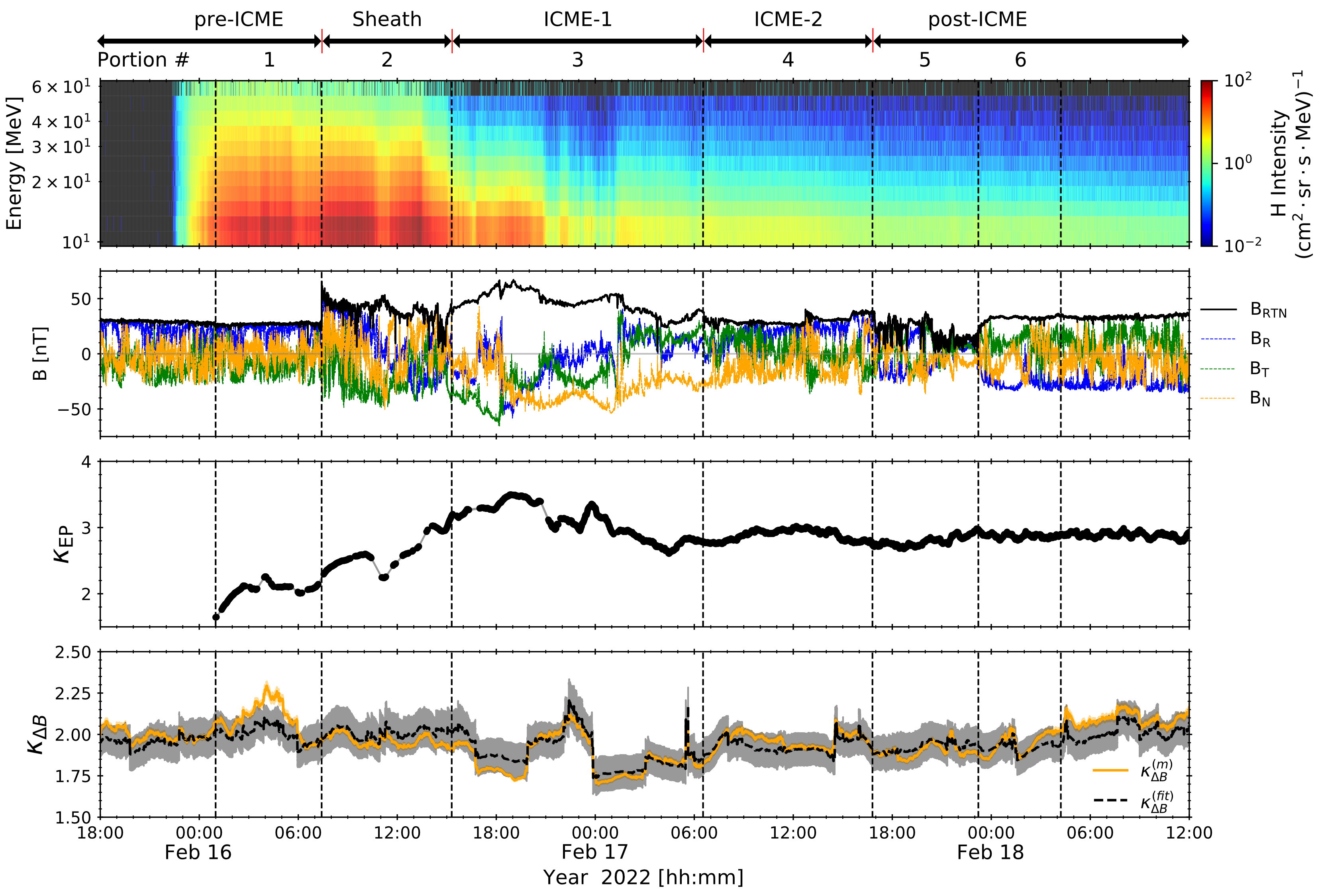}
    \caption{The overview of the SEP event observed by Parker Solar Probe on 2022 February 15. In descending order, the panels include the spectrogram of energetic protons observed by HET-A, the magnetic field (RTN components in blue, green and yellow, respectively, and its magnitude in black), the thermodynamic kappa of energetic protons observed between 10--60~MeV, and the overall intermittent kappa via two methods: ratio of moments method (solid yellow line) and fitting the distribution of magnetic increments (dashed black line), each surrounded by a shadow representing their error corresponding to one standard deviation.}
    \label{fig:Overview}
\end{figure}

On 2022 February 15
-18
, PSP observed an SEP event associated with an ICME, among other spacecraft \citep{PalmerioEA2024ApJ_CME_BepiPSP_2022feb15,KhooEA2024ApJ_CME_BepiPSP_2022feb15}.
We provide an overview of PSP observations in Figure \ref{fig:Overview}, including a spectrogram of energetic proton intensities observed by HET-A onboard the IS\(\odot\)IS instrument suite and the magnetic field in RTN coordinates.
Values of the thermodynamic \(\kappa_{\rm EP}\) for the energetic protons are given, in addition to values of the overall intermittent kappa derived in two ways: (1) fitting the PDF of unsigned magnetic increments using Eq. \eqref{eq:fit_kappa} leading to \(\kappa_{\Delta B}^{\rm (fit)}\) and (2) solving for \(\kappa_{\Delta B}^{\rm (m)}\) via Eq. \eqref{eq:powermean_ratio} by computing the ratio of power-means empirically.
We compute a time series of \(\kappa_{\Delta B}\) derived from each method by computing a single value from a 3~hour, overlapping interval that is shifted every 1~second.
Values of \(\kappa_{\rm EP}\) begin slightly later than the onset time of observed particles due to criteria that determines the stationarity of the window used to compute \(\kappa_{\rm EP}\) \citep[][see Section 2]{CuestaEA2024ApJ_EPthermoKappaObservations}.
Therefore, since the proton intensity distribution is changing significantly for some time after the event onset time, many of the resulting values of \(\kappa_{\rm EP}\) are discarded.

\newpage

\begin{figure}[ht]
    \centering
    \includegraphics[width=\linewidth]{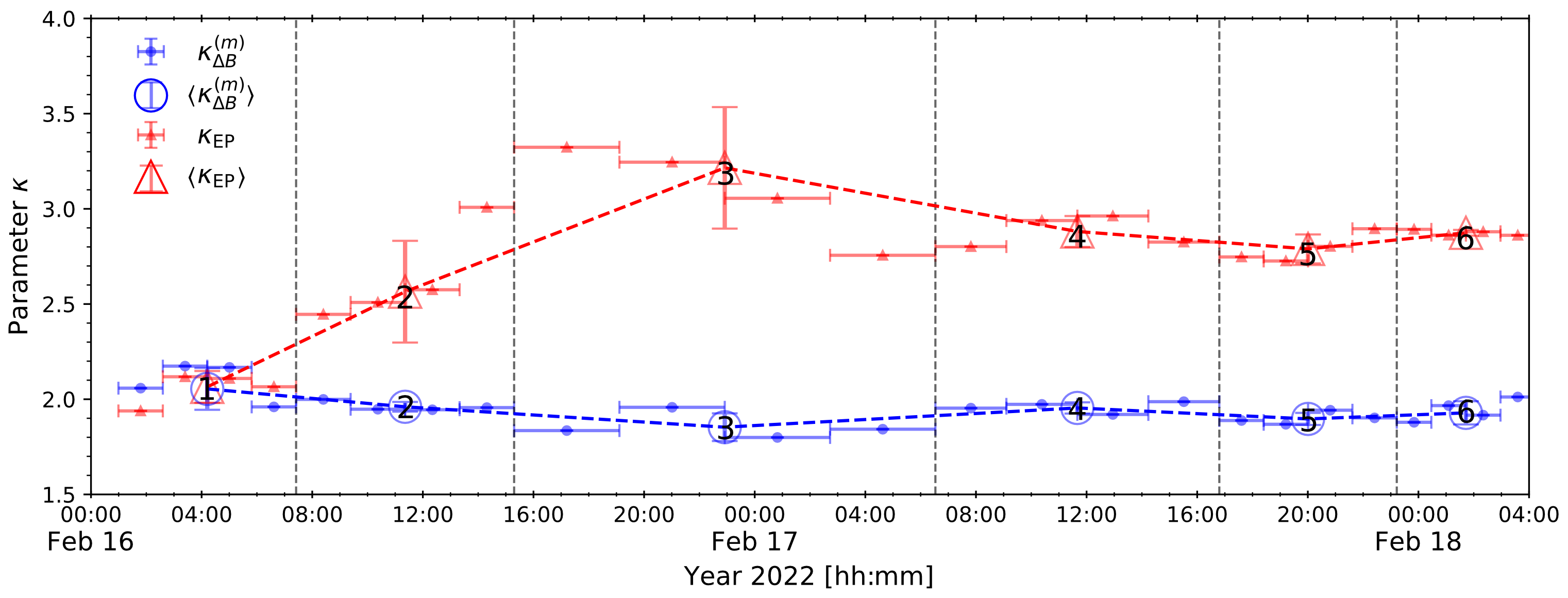}
    \includegraphics[width=\textwidth]{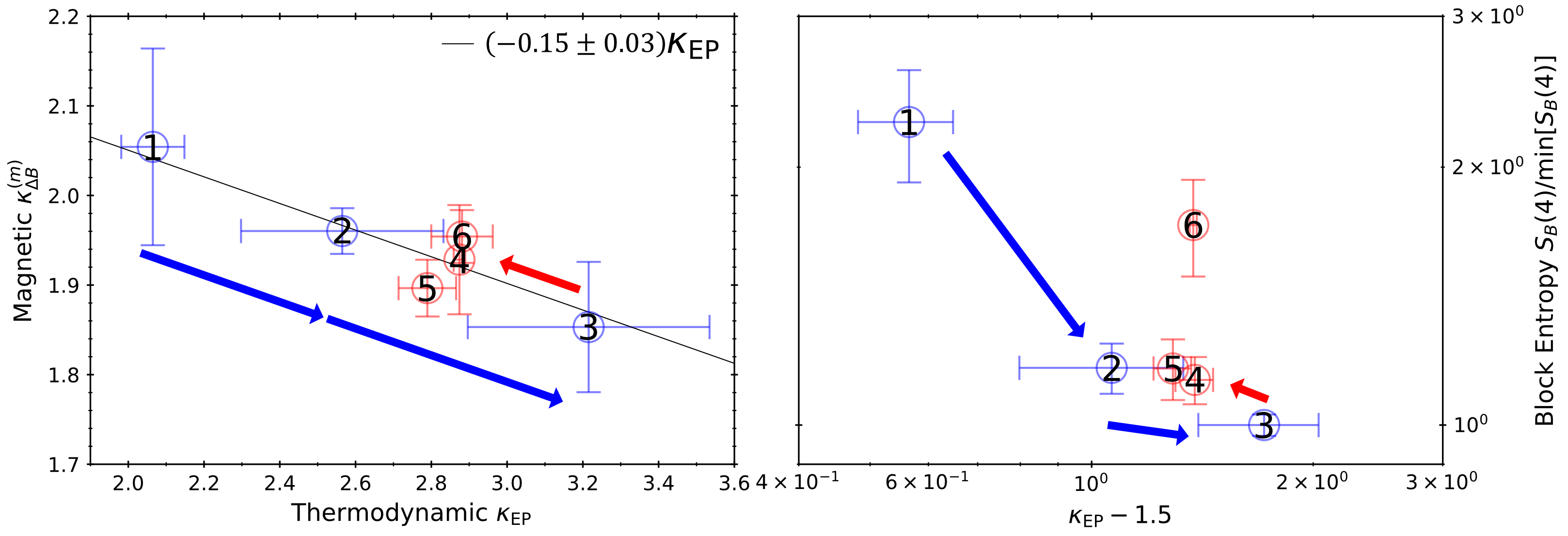}
    \caption{Top Panel: The time series of \(\kappa_{\rm EP}\) (red triangles) and \(\kappa_{\Delta B}^{\rm (m)}\) (blue circles) for the 2022 February 15 SEP event. Bottom left panel: The time series from the top panel collapsed over time, showing the exchange of correlation between the magnetic field and energetic protons. Bottom right panel: The block entropy as a function of the thermodynamic \(\kappa_0=\kappa_{\rm EP}-3/2\), which is related to the thermodynamic entropy for the energetic protons. The numbers within their respective markers represent the portion number of the ICME, which is marked in Figure \ref{fig:Overview}.}
    \label{fig:GroupedEvolution}
\end{figure}

We divide the SEP event into a total of six portions marking different regions of the ICME.
The start and end times of these portions, along with their identification, were taken from \citet{PalmerioEA2024ApJ_CME_BepiPSP_2022feb15}.
Note that the ICME-1 and ICME-2 portions mark different regions of the same ICME, and do not represent two separate ICMEs.
To find the overall evolution of \(\kappa_{\Delta B}\) and \(\kappa_{\rm EP}\), we group these time series by these six portions and take a series of weighted-averages.
We have shown in Figure \ref{fig:Overview} that \(\kappa_{\Delta B}^{\rm (m)}\) is consistent with \(\kappa_{\Delta B}^{\rm (fit)}\) within their errors and has consistently smaller error; therefore, we proceed with comparing only \(\kappa_{\Delta B}^{\rm (m)}\) with \(\kappa_{\rm EP}\).
The first averaging procedure involves dividing each portion into four equal parts and computing a weighted average over each equal part.
Second, we compute another weighted average of the four resulting values resulting in one average value of \(\kappa_{\Delta B}^{\rm (m)}\) and \(\kappa_{\rm EP}\) per portion.
These results are shown in the top panel of Figure \ref{fig:GroupedEvolution}.

The evolution of \(\kappa_{\rm EP}\) throughout the portions of the ICME may be explained via the gain and loss of correlations among the observed SEPs.
For example, as \(\kappa_{\rm EP}\) increases from the pre-ICME region into the ICME-1 region, the particles are losing correlation \citep[since \(\kappa \sim \rho^-1\) for level of correlation \(\rho\),][]{LivadiotisMcComas2010ApJ_SpacePlasmaNonEquil,LivadiotisMcComas2013SSRv_KappaDist_SpSc}.
Then from regions ICME-1 to post-ICME, the particles gain correlations, which effectively reduces \(\kappa_{\rm EP}\).
Similar reasoning applies for \(\kappa_{\Delta B}\), such that it first gains correlations (decreasing \(\kappa\)) from the pre-ICME region into the ICME-1 region, then it loses correlations (increasing \(\kappa\)) from ICME-1 to the post-ICME region.
A potential mechanism responsible for this behavior is discussed further in Section \ref{sec:discussion}.

Next, we collapse their weighted averages over time and plot them as a function of each other, which is shown in the bottom left panel of Figure \ref{fig:GroupedEvolution}.
The bottom right panel shows the block entropy as a function of \(\kappa_0=\kappa_{\rm EP}-3/2\), which is analogous to the thermodynamic entropy.
In both bottom panels, we color the first three portions blue indicating a decreasing (increasing) entropy in the magnetic field (energetic protons), and the following portions are colored red indicating a reversal in this behavior.
As a result, the bottom panels of Figure \ref{fig:GroupedEvolution} reveal that variations between \(\kappa_{\Delta B}\) and \(\kappa_{\rm EP}\) are described by a linear relationship.
This suggests that a transfer of entropy from the magnetic field to energetic protons occurs.


\newpage

\newpage
\section{Discussion} \label{sec:discussion}

The levels of correlation for fluctuations in the magnetic field and in energetic particle thermal velocities are related to \(\kappa_{\Delta B}\) and \(\kappa_{\rm EP}\), respectively.
A higher kappa value signifies a lower degree of correlation, and vice versa \citep{LivadiotisMcComas2010ApJ_SpacePlasmaNonEquil,LivadiotisMcComas2013SSRv_KappaDist_SpSc}.
Despite being different quantities measured and calculated independently, Figure~\ref{fig:GroupedEvolution} reveals a correlation between \(\kappa_{\Delta B}\) and \(\kappa_{\rm EP}\).
We also examine the block entropy (\(S_B\)), a type of information entropy, of the magnetic field magnitude, which contains information of sequences and their probability of occurrence in the measurements \citep{Grassberger1989IEEE_BlockEntropy,Livadiotis2008PhyA_BlockEntropy}.
We find that \(S_B\) follows a similar behavior of decreasing and increasing trends with respect to \(\kappa_{\rm EP}-3/2\), which is connected to the thermodynamic entropy.

The observed correlation between \(\kappa_{\Delta B}\) and \(\kappa_{\rm EP}\) may be more appropriately viewed in terms of a transfer of degrees of freedom, because \(\kappa\) is related to the effective degrees of freedom (\({\rm d_{eff}}\)) as $\kappa = A + {{\rm d_{eff}}}/{b}$, 
where \(A\) and \(b\) are constants \citep{Livadiotis2017kappa_book,LivadiotisEA2024ApJ_PUIthermo}.
Unlike regular degrees of freedom, \({\rm d_{eff}}\) accounts for the fact that not all degrees of freedom contribute equally.
Since there may be inactive modes that do not contribute to a system's thermodynamic properties, \({\rm d_{eff}}\) can only be equal to or less than the regular degrees of freedom.
Therefore, changes in \(\kappa\) are coupled to changes in \({\rm d_{eff}}\), with the level of coupling depending on $1/b$. We define $1/b$ as the coupling coefficient $C$, so that:

\begin{equation}\label{eq:kappa_dof}
    \kappa = A + C \cdot {\rm d_{eff}}\text{.}
\end{equation}
If the magnetic field loses degrees of freedom, they may be transferred to another system, such as energetic particles. By combining Eq. \eqref{eq:kappa_dof} for both systems, the transfer of ${\rm \Delta d}$ degrees of freedom can be described as:

\begin{equation}\label{eq:dof_mag}
    \kappa_{\Delta B} = A_{\Delta B} + C_{\Delta B} \cdot \left( {\rm d}_{\rm eff}^{(\Delta B)} - {\rm \Delta d} \right)
\end{equation}
and 

\begin{equation}\label{eq:dof_ep}
    \kappa_{\rm EP} = A_{\rm EP} + C_{\rm EP} \cdot \left( {\rm d}^{\rm (EP)}_{\rm eff} + {\rm \Delta d} \right)
\end{equation}
since a gain/loss of \({\rm d}_{\rm eff}^{(\Delta B)}\) corresponds to a loss/gain of \({\rm d^{({\rm EP})}_{\rm eff}}\).
By solving Eqs. \eqref{eq:dof_mag} and \eqref{eq:dof_ep} for their corresponding \( {\rm \Delta d}\), we find:

\begin{equation}\label{eq:linear_kappadof}
    -\left[ C^{-1}_{\Delta B}\cdot (\kappa_{\Delta B}-A_{\Delta B})-{\rm d_{eff}^{(\Delta B)}}\right] = C^{-1}_{\rm EP}\cdot (\kappa_{\rm EP}-A_{\rm EP})-{\rm d_{eff}^{(EP)}} = {\rm \Delta d}\text{.}
\end{equation}
Eq. \eqref{eq:linear_kappadof} implies that there is a linear relationship between \(\kappa_{\Delta B}\) and \(\kappa_{\rm EP}\) describing the transfer of \({\rm d_{\rm eff}}\), given by:

\begin{equation}\label{eq:kappa_coupling}
    \kappa_{\Delta B}={\rm Intercept}- \left( \frac{C_{\Delta B}}{C_{\rm EP}} \right) \kappa_{\rm EP}
\end{equation}
with \({\rm Intercept} = \left( {\rm d_{eff}^{(EP)}} + {\rm d_{eff}}^{(\Delta B)} + A_{\Delta B}C^{-1}_{\Delta B} - A_{\rm EP}C^{-1}_{\rm EP}\right)\cdot C_{\Delta B}\). If Intercept is constant, then Eq. \eqref{eq:kappa_coupling} is a linear relationship between \(\kappa_{\Delta B}\) and \(\kappa_{\rm EP}\). 
Since the intercept contains ${\rm d_{eff}^{(EP)}} + {\rm d_{eff}}^{(\Delta B)}$,
and the degrees of freedom are not constant, the intercept does not have to be constant. 
But we observe a linear relationship between \(\kappa_{\Delta B}\) and \(\kappa_{\rm EP}\), given by \(\kappa_{\Delta B} \sim (-0.15 \pm 0.03)\kappa_{\rm EP}\). $A$ and $b$ are constants, so we can infer that ${\rm d_{eff}^{(EP)}} + {\rm d_{eff}}^{(\Delta B)}$ is constant, implying that the total degrees of freedom is conserved. This could mean that degrees of freedom were transferred from the magnetic field to the energetic particles.
The coefficient of the linear relationship implies that the coupling coefficient ratio is \(C_{\Delta B}/C_{\rm EP}\sim 0.15\pm0.03\). Physically, this means that \(\kappa_{\Delta B}\) is less influenced by changes in \({\rm d_{eff}}\) than \(\kappa_{\rm EP}\).

The correlation between \(\kappa_{\Delta B}\) and \(\kappa_{\rm EP}\) can also be explained in terms of the entropic rate of change (\(\dot{\sigma}\)), given by:

\begin{equation} \label{eq:entropy_transport}
    \frac{d\left( \frac{1}{\kappa}\right)}{dt} = -\frac{2\dot{\sigma}}{S^2_\infty}
\end{equation}
where \(S_\infty\) represents the BG entropy per particle at classical thermal equilibrium \citep{LivadiotisMcComas2021Entrp_TemperatureKappa_Entropy,LivadiotisMcComas2023ApJ_TransportEquation_Kappa}.
The entropy transport Eq. \eqref{eq:entropy_transport} can be physically understood through the concept of entropy defect \citep{LivadiotisMcComas2021Entrp_TemperatureKappa_Entropy,LivadiotisMcComas2022ApJ_KappaPhysicalCorrelations,LivadiotisMcComas2023ApJ_TransportEquation_Kappa,LivadiotisMcComas2023NatSR_EntropyDefect,LivadiotisMcComas2023EPL_EntropyDefect_Algebra,LivadiotisMcComas2024EPL_KappaUniversality}.
Entropy defect measures how much the entropy changes.
The entropy of the combined system of particles is less than the sum of the entropies of the subsystems, which is due to long-range correlations between the particles.
Therefore, entropy defect measures the missing entropy between the system and the sum of the individual entropies of its parts.
According to Eq. \eqref{eq:entropy_transport}, a decreasing \(\kappa\) (\({\rm d_{eff}}-{\rm \Delta d}\)) leads to a negative entropy rate \(\dot{\sigma}<0\); and vice versa, namely, an increasing \(\kappa\) (\({\rm d_{eff}}+{\rm \Delta d}\)) leads to a positive entropy rate \(\dot{\sigma}>0\).
Consequentially, the corresponding change in entropy of the energetic particles must have been transferred from another system, such as the magnetic field.

This raises the question: What causes the relationship between \(\kappa_{\Delta B}\) and \(\kappa_{\rm EP}\)?
Magnetic fields directly influence energetic particles through the Lorentz force, so it is plausible that changes in the magnetic field fluctuations can affect the level of correlation between energetic particles.
Due to their low density, it is unlikely that energetic particles can affect the magnetic field significantly like the solar wind plasma does by dragging field lines due to the frozen-in condition.
Therefore, the causal factor is likely with the magnetic field, which affects the energetic particles as a byproduct.
However, we note that under some circumstances, energetic particles may influence the dynamics of the magnetic field as observed during particle energization induced by flux-rope deformation and merging \citep{DrakeEA2006NatAcceleration,OkaEA2010ApJ_ElectronAccel_Coalescence,DrakeEA2013ApJL_EnergeticParticleSpectra_Reconnection,PhanEA2024ApJL_MagReconnection_HCS_FluxRopeMerging,DesaiEA2024arXiv_HCSMagReconnection_ProtonEnergization}.

\begin{figure}[ht]
    \centering
    \includegraphics[width=\linewidth]{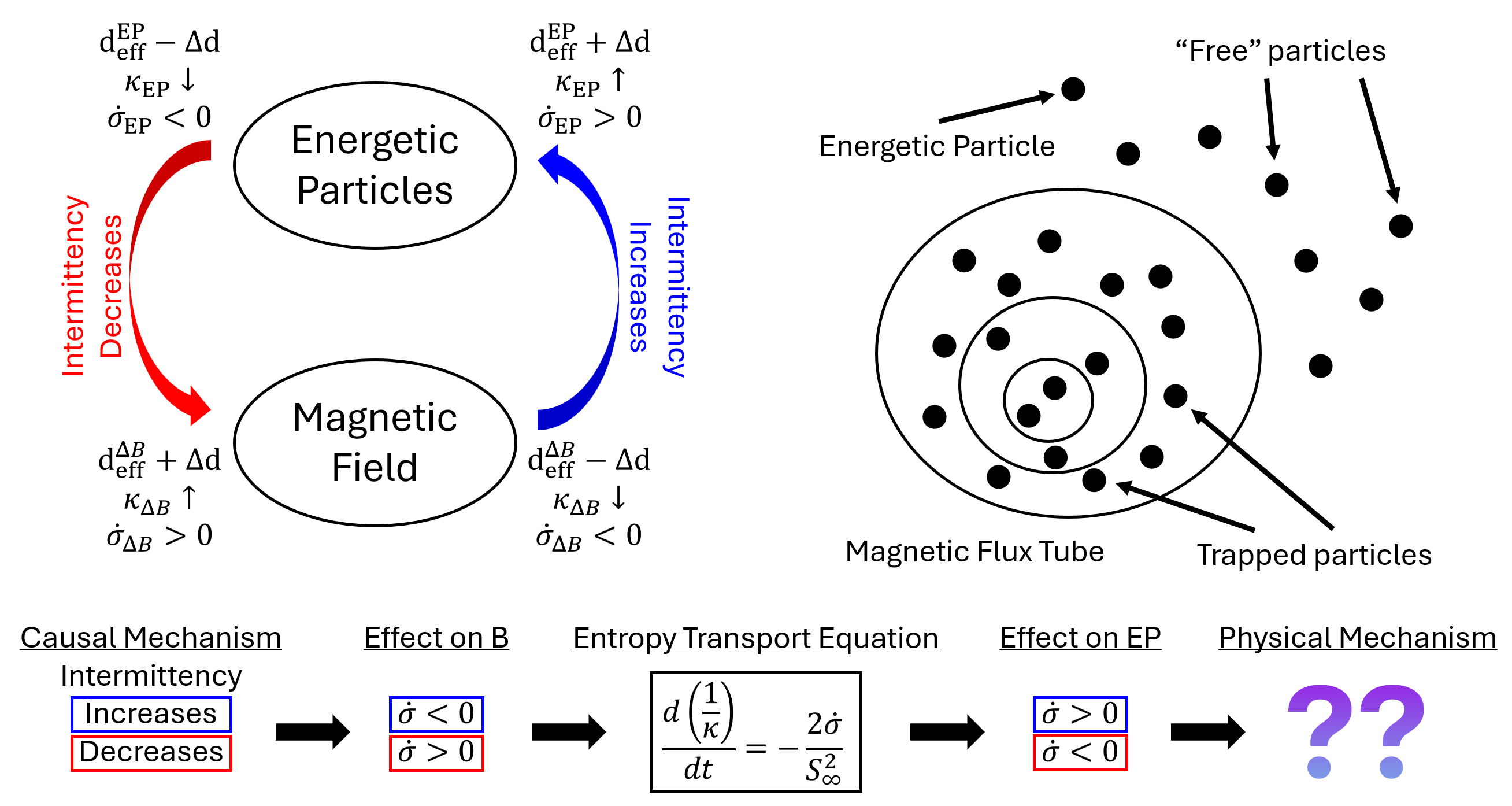}
    \caption{An illustration of the transfer of entropy, or degrees of freedom, from the magnetic field to energetic particles. Also depicted are energetic particles trapped and modulated by a magnetic flux tube which may possibly be a thermal mechanism of the transport of entropy affecting the energetic particles. The center boxed equation describes the transport of entropy over time \citep{LivadiotisMcComas2023ApJ_TransportEquation_Kappa}.}
    \label{fig:illustration}
\end{figure}

A possible explanation for our findings is that the energetic particles, through some unknown process(es), acts as a reservoir for the magnetic field entropy, as illustrated in Figure \ref{fig:illustration}.
When coherent structures become more dominant over localized fluctuations, the intermittency in the magnetic field (the causal mechanism) increases, decreasing \(\kappa_{\Delta B}\), resulting in (\(\dot{\sigma}<0\)).
Such coherent structures can accelerate
\citep{TesseinEA2013ApJL_PVI_intensity,TesseinEA2015ApJ_EPs_PVI,KhabarovaEA2015ApJ_MagIslands_acceleration,KhabarovaZank2017ApJ_EPs_CurrentSheets,MalandrakiEA2019ApJ_EPs_CurrentSheets,BandyopadhyayEA2020ApJS_PVI_intensity,CuestaEA2024ApJ_CMEShockCorrelation}
and modulate 
\citep{TesseinEA2015ApJ_EPs_PVI,Tessein2016GRL_EPtrapping_CoherentStructures,PecoraEA2021MNRAS_SEP_helicity} energetic particles.
Because there are known mechanisms through which the magnetic field strongly influences the energetic particles,
it's plausible that increasing the intermittency of the magnetic field, which reduces the entropy of the magnetic field, transfers entropy to the energetic particles. For example, acceleration of lower energy particles into the SEP range may add entropy to the SEP system and decrease correlations between the SEPs, increasing \(\kappa_{\rm EP}\).
For this SEP event, we were unable to test this theory due to the lack of high-resolution plasma data during this period, which is needed for the rigorous detection of coherent structures such as small-scale flux ropes, an important class of coherent structures that can affect SEPs \citep{FarookiEA2024ApJ_FluxRopeDetection}.
The modulation of SEPs by coherent structures could explain changes in \(\kappa_{\rm EP}\) in response to \(\kappa_{\Delta B}\) because high energy SEPs are more likely to escape coherent structures than lower energy SEPs.
As a result, the spectral index \citep[equivalent to \(\kappa_{\rm EP}\);][]{LivadiotisEA2024ApJ_EPthermoKappaTechnique,CuestaEA2024ApJ_EPthermoKappaObservations} would increase.

\section{Conclusions} \label{sec:summary}

Recent studies have shown that thermodynamic properties of energetic particles can be derived from observations of their energy distribution in the high-energy limit \citep{LivadiotisEA2024ApJ_EPthermoKappaTechnique,CuestaEA2024ApJ_EPthermoKappaObservations}.
These high energy tails are used to fit the kappa distribution formula, which is connected to thermodynamics in the presence of long-range correlations among the electromagnetic field and particles \citep{LivadiotisMcComas2012ApJ_KappaThermoProcesses}.
Consequently, the thermodynamic kappa for energetic particles (\(\kappa_{\rm EP}\)) holds information regarding the effective degrees of freedom (\({\rm d_{eff}}\)) among the particle population, in addition to their level of correlation, which has an inverse relationship with \(\kappa\) \citep{LivadiotisMcComas2013SSRv_KappaDist_SpSc}.

The distribution of magnetic increments has also been frequently described as having a kappa-like behavior \citep{GrecoEA2018}.
Therefore, a parameter kappa for magnetic increments (\(\kappa_{\Delta B}\)) can also be derived from magnetic field observations.
We investigated two methods for deriving \(\kappa_{\Delta B}\) including (1) fitting the probability density function of unsigned magnetic increments via Eq. \eqref{eq:fit_kappa} (yielding values of \(\kappa^{\rm (fit)}_{\Delta B}\)) and (2) empirically finding the ratio of moments and solving for kappa via Eq. \eqref{eq:powermean_ratio} (yielding values of \(\kappa^{\rm (m)}_{\Delta B}\)).
These two methods are demonstrated in Figure \ref{fig:SampleMethod}. In Figure \ref{fig:Overview}, we show that these two methods of finding \(\kappa_{\Delta B}\) are consistently equal within their fitting errors over the entire SEP event.
Therefore, we use \(\kappa_{\Delta B}^{\rm (m)}\) to compare with \(\kappa_{\rm EP}\) as a function of time over different portions of the ICME observed on 2022 February 16 \citep{PalmerioEA2024ApJ_CME_BepiPSP_2022feb15,KhooEA2024ApJ_CME_BepiPSP_2022feb15}.

Finally, we explore the relationship between \(\kappa_{\Delta B}\) and \(\kappa_{\rm EP}\) during the passing of an ICME.
This relationship can be described as a transfer of entropy from the magnetic field to energetic protons in the form of \({\rm d_{eff}}\), which is shown in the bottom panels of Figure \ref{fig:GroupedEvolution}.
Leading into the ICME from portion 1 to portions 2 and 3, \(\kappa_{\Delta B}\) decreases as \(\kappa_{\rm EP}\) increases.
Then the relationship flips, such that \(\kappa_{\Delta B}\) increases as \(\kappa_{\rm EP}\) decreases.
We observed that \(\kappa_{\Delta B}\sim (-0.15 \pm 0.03)\kappa_{\rm EP}\), exhibiting an anti-correlated relationship.
This signifies that the coupling effectiveness between the energetic particles' correlations and their \({\rm d_{eff}}\) is stronger than the magnetic field.
Furthermore, the transfer of entropy over the ICME does not make a complete cycle since the energetic particles end up with a larger value of \(\kappa_{\rm EP}\) towards the end of the SEP event.
Thus, the ICME provides a complex environment where the transfer of entropy between the magnetic field and energetic particles can occur without loss in their combined \({\rm d_{eff}}\).

In a broader context, these results provide a new perspective on the thermodynamic interactions between a magnetic field and energetic particles traveling through it.
These findings are fundamentally important for understanding the complex dynamics of ICMEs and their impact on the thermodynamic conditions of interplanetary space.
The anti-correlation between \(\kappa_{\Delta B}\) and \(\kappa_{\rm EP}\) may suggest a deeper connection between magnetic field fluctuations and particle behavior, which could exist in heliospheric plasmas and even astrophysical plasmas beyond the heliosphere.

\section{Acknowledgments}

We thank the IS\(\odot\)IS team and everyone that made the PSP mission possible. 
The IS\(\odot\)IS data and visualization tools are available to the community at \href{https://spacephysics.princeton.edu/missions-instruments/PSP}{https://spacephysics.princeton.edu/missions-instruments/PSP}. 
PSP was designed, built, and is operated by the Johns Hopkins Applied Physics Laboratory as part of NASA’s Living with a Star (LWS) program (contract NNN06AA01C).


\newpage
    
\bibliographystyle{plainnat}





\end{document}